\documentclass[reqno,12pt]{amsart}

\usepackage{amssymb,graphicx}
\usepackage{verbatim}

\theoremstyle{plain}
\newtheorem{test}{Definition}[section]
\newtheorem{sss}[test]{Definition}
\newtheorem{Reynoldsequation}[test]{Proposition}
\newtheorem{grashoflemma}{Lemma}[section]

\newtheorem{Glimit}[grashoflemma]{Definition}
\newtheorem{HB}{Proposition}[section]
\newtheorem{galerkapprox}[HB]{Proposition}
\newtheorem{thm1}{Theorem}[section]
\newtheorem{thm2}[thm1]{Theorem}
\newtheorem{enupper}[thm1]{Proposition}

\newtheorem{CDestimate}[thm1]{Proposition}
\newtheorem{corCD}[thm1]{Theorem}

\newtheorem{corCD2}[thm1]{Proposition}

\newtheorem{cor2}[thm1]{Theorem}

\newtheorem{coefupper}[thm1]{Corollary}
\newtheorem{deltadefinition}{Definition}[section]

\newtheorem{upperkappa}[deltadefinition]{Proposition}

\theoremstyle{definition}
\newtheorem{rmk}{Remark}[section]

\newtheorem{kenergyP}{Proposition}[section]

\newtheorem{componentestimate}{Lemma}[section]
\newtheorem{prop}{Proposition}[section]

\newcommand{\NN}{\mathbb{N}}
\newcommand{\ZZ}{\mathbb{Z}}
\newcommand{\RR}{\mathbb{R}}

\newcommand{\bu}{\mathbf{u}}
\newcommand{\bv}{\mathbf{v}}
\newcommand{\bw}{\mathbf{w}}
\newcommand{\be}{\mathbf{e}}

\newcommand{\bx}{\mathbf{x}}

\newcommand{\bbf}{\mathbf{f}}

\newcommand{\bU}{\mathbf{U}}

\newcommand{\bnabla}{\boldsymbol{\nabla}}

\newcommand{\loc}{{\text{\rm loc}}}

\newcommand{\Ucal}{{\mathcal U}}

\newcommand{\Rey}{{\text{\rm Re}}}

\newcommand{\rd}{{\text{\rm d}}}
\newcommand{\rw}{{\text{\rm w}}}

\newcommand{\ddt}[1]{\frac{\text{\rm d}#1}{\text{\rm dt}}}

\newcommand{\average}[1]{\langle{#1}\rangle}

\newcommand{\Vinner}[1]{(\!({#1})\!)}
\newcommand{\Lim}{\operatorname*{\textsc{Lim}}_{T\rightarrow \infty}}

\begin{document}
\numberwithin{equation}{section}


\title{Statistical estimates for channel flows driven by a pressure gradient}

\author{F. Ramos${}^1$}
\author{R. Rosa${}^1$}
\author{R. Temam${}^{2,3}$}

\address{${}^1$ Departamento de Matem\'atica Aplicada \\
  Instituto de Matem\'atica \\ Universidade Federal do Rio de Janeiro \\
  Caixa Postal 68530 Ilha do Fund\~ao \\ Rio de Janeiro RJ 21945-970 \\
  Brazil}
\address{${}^2$ Department of Mathematics, Indiana University, Bloomington,
  IN 47405}
\address{${}^{3}$ Laboratoire d'analyse num\'erique et EDP,
  Universit\'e de Paris-Sud, Orsay, France}

\email[F. Ramos]{framos@ufrj.br}
\email[R. Rosa]{rrosa@ufrj.br}
\email[R. Temam]{temam@indiana.edu}

\thanks{This work was partly supported by the Pronex in Turbulence, CNPq
   and FAPERJ, Brazil, grant number E-26/171.198/2003, by NSF grant
      number DMS-0305110, and by CNPq, Bras\'{\i}lia, Brazil, grant number
      30.0902/2003-4.}

\date{December 14, 2006}


\begin{abstract}
We present rigorous estimates for some physical quantities related
to turbulent and non-turbulent channel flows driven by a uniform
pressure gradient. Such results are based on the concept of
stationary statistical solution, which is related to the notion of
ensemble average for flows in statistical equilibrium. We provide a
lower bound estimate for the mean skin friction coefficient and
improve on a previous upper bound estimate for the same quantity;
both estimates in terms of the Reynolds number. We also present
lower and upper bound estimates for the mean rate of energy
dissipation, the mean longitudinal velocity (in the direction of the
pressure gradient), and the mean kinetic energy. In particular, we
obtain an upper bound related to the energy dissipation law, namely
that the mean rate of energy dissipation is essentially bounded by a
non-dimensional universal constant times the cube of the mean
longitudinal velocity over a characteristic macro-scale length.
Finally, we investigate the scale-by-scale energy injection due to
the pressure gradient, proving an upper bound estimate for the
decrease of this energy injection as the scale length decreases.
\end{abstract}

\maketitle

\section{Introduction}

While the existence of exact solutions of the Navier-Stokes Equations
are not available in general, most of the classical research on
turbulence theory consist of approximate methods based on a few
exact deductions, supplemented with intuitive hypotheses
about the nature of the phenomenon, such as scaling assumptions and
moment truncation models; see for example \cite{batchelor, TL}.

Recently, part of the theoretical research on turbulence has been
concentrated on deriving rigorous bounds on characteristic quantities
of turbulent flows directly from the equations of motion.
These results are important to substantiate the ones obtained via the classical approximation methods.

Decomposing the turbulent flow into a stationary background flow and a
fluctuation component, and using variational methods, Constantin and
Doering derived rigorous results for the long-time averaged rate of energy
dissipation of flows in some geometries, in particular for the channel flow
driven by a pressure gradient, in which case the estimate also yields an
estimate for the mean skin friction coefficient;
see \cite{constantindoering94, constantindoering95}.

Meanwhile, rigorous results were recently established for the
three-dimensional theory of homogeneous stationary statistical
turbulence in \cite{fmrt, fmrtcr1, fmrtcr2} using the concepts of
stationary statistical solutions of the Navier-Stokes equations and
generalized time average measures, and using energy-type methods.

This paper presents a combination of those results in the specific
case of channel flows driven by a uniform pressure gradient. More
specifically, we extend the upper bound estimate for the long-time
averaged rate skin friction coefficient, obtained in
\cite{constantindoering95}, to general stationary statistical
solutions, slightly simplifying their proof and slightly improving
their estimates. We also obtain a lower bound estimate for the skin
friction coefficient, which cannot be obtained by the variational
principle method of \cite{constantindoering95}. More precisely, we
show that for every stationary statistical solution the
corresponding skin friction coefficient $C_f$ satisfies
\[ \frac{10.88}{\Rey}
       \leq C_f \leq \frac{13.5}{\Rey},
\]
for low Reynolds number flows, and
\[ \frac{10.88}{\Rey}
     \leq C_f \leq 0.484 + O\left(\frac{1}{\Rey}\right),
\]
for high Reynolds number flows, 
where the Reynolds number is defined by $\Rey=hU/\nu$, with
$h$ being the height of the channel and $U$, the mean longitudinal
velocity. 

The lower-bound estimate for $C_f$ is nearly optimal in the sense that the
stationary statistical solution is arbitrary and may be concentrated
on the plane Poiseuille flow (which is unstable for high Reynolds number
flows, but anyway exists in a mathematical sense), 
for which $C_f=12/\Rey$. The
upper-bound estimate might not be optimal since heuristic arguments
suggest that $C_f\sim (\ln \Rey)^{-2}$ for high-Reynolds number
turbulent flows. Nevertheless, it represents a nearly 19\%
improvement over the estimate obtained in \cite{constantindoering95}
on the leading order constant term (from $0.597$ to $0.484$.)

We also give upper and lower bound estimates for some other physical
quantities, such as the mean energy dissipation rate, the mean
kinetic energy, and the mean longitudinal velocity. In particular,
we prove an upper bound estimate related to the energy dissipation
law, namely that for large Reynolds number flows the mean rate of 
energy dissipation is essentially bounded by a non-dimensional universal 
constant times the cube of the mean
longitudinal velocity $U$ over the height $h$ of the channel:
\[ \epsilon \leq \left( 0.054 + O\left(\frac{1}{\Rey^2}\right)
         \right)\frac{U^3}{h}.
\]
The leading order constant term obtained in \cite{constantindoering95}
was approximately $0.0884$, and it was remarked in that work that
this term is much lower than $1$, hinting that this result 
is substantially more than a formalized dimensional analysis argument.
The same applies here.

Finally, we study the scale-by-scale energy injection term due to
the pressure gradient. We show that the energy injected into the
modes larger than or equal to $\kappa$ is bounded
by a term proportional to $\kappa^{-3/2}$. The motivation for the
study of the decrease of energy injection comes from the Kolmogorov
theory of turbulence. This theory argues that for turbulent flows
there is a certain range of scales much lower than the energy
injection scales and greater than the energy dissipative scales in
which the kinetic energy is transferred to the small scales at a
nearly constant rate equal to the energy dissipation rate. This
theory was proposed in the idealized case of locally homogeneous
turbulence, away from the boundaries, under the assumption that the
energy injection is concentrated on the large scales. However, it is
known from experiments that for wall bounded turbulence this
hypothesis needs to be corrected \cite{constantindoering95, TL,
Prandtl}. In particular, the energy injection occurs at arbitrarily
small scales. The estimates presented yield an upper bound on the
rate of decrease of energy injection as the scale length decreases.

The remaining of the paper is organized as follows. In the next section
we introduce the convenient mathematical setting used throughout the
work. In Section $3$, we define the notion of stationary statistical
solution of the Navier-Stokes equations and present some related
results. In Section $4$, we rigorously define the characteristic
quantities that will be estimated, such as the mean energy
dissipation rate, mean kinetic energy, mean longitudinal velocity,
and mean skin friction coefficient. In Section $5$, we establish a
relation between stationary statistical solutions and time averages.
In Sections $6$ and $7$, we explicitly derive rigorous bounds for
the mentioned physical quantities, utilizing both methods of
\cite{fmrt, fmrtcr1, fmrtcr2} and of \cite{constantindoering95}. In
Section $8$, we list the explicit values of the mentioned physical
quantities for the specific case of the laminar Poiseuille flow,
verifying that some of the results obtained in Section $6$ and
Section $7$ are optimal in some sense. In Section $9$, we conclude
the work with the discussion about the scale-by-scale energy
injection.

\section{Mathematical framework of the Navier-Stokes Equations}

We consider an incompressible Newtonian flow confined to a
rectangular periodic channel and driven by a uniform pressure gradient.
More precisely, the velocity
vector field $\bu=(u_1,u_2,u_3)$ of the fluid satisfies the incompressible
Navier-Stokes equations
\begin{equation}
\label{nseeq1}
\frac{\partial \bu}{\partial t} - \nu \Delta \bu
       + (\bu\cdot\bnabla)\bu + \bnabla p = \frac{P}{L_{x}}\mathbf{e}_{1}, 
    \qquad
      \bnabla \cdot \bu = 0,
\end{equation}
in the domain
$\Omega=\left(0,L_{x}\right)\times\left(0,L_{y}\right)\times\left(0,h\right)$.
The scalar $p$ is the kinematic pressure. We denote by
$\bx=\left(x,y,z\right)$ the space variable.
The boundary conditions are no-slip on the planes $z=0$ and $z=h$
and periodic in the $x$ and $y$ directions, with periods $L_{x}$
and $L_{y}$, respectively, for both $\bu$ and $p$. The parameter
$P/L_{x}$ denotes the magnitude of the applied pressure gradient. 
The parameter $\nu>0$ is the
kinematic viscosity, $\mathbf{e}_{1}$ is the unit vector in the
$x$ direction, and $L_{x}$, $L_{y}$, $h$, $P>0.$ We sometimes refer to 
the direction $x$ of the pressure gradient as the longitudinal direction.

The mathematical formulation of the Navier-Stokes equations in this geometry
can be easily adapted from the no-slip or fully-periodic case developed in
\cite{constantinfoias, fmrt, lady63, temam, temam3}.

The  formulation yields a functional equation for the time-dependent velocity field
$\bu=\bu(t)$ of the form:
\begin{equation}
  \label{nseeq}
  \ddt{\bu} = \mathbf{F}(\bu) = \mathbf{f}_{P} - \nu A\bu - B(\bu,\bu),
\end{equation}
where
\begin{equation}
  \bbf_P=\frac{P}{L_x}\be_1.
\end{equation}
Two fundamental spaces are defined by
\[ H=\left\{ \bu=\bw|_{\Omega};\;
      \begin{aligned}
      &  \bw\in  (L^{2}_\loc(\RR^2\times(0,h)))^{3}, \;\bnabla\cdot\bw=0,
      \;\\
         &\bw(x+L_{x},y,z)=\bw(x,y,z),\\
      &  \bw(x,y+L_{y},z)=\bw(x,y,z), \text{ a.e. } (x,y,z)\in \mathbb{R}^2\times (0,h).\\
      &  \bw_{3}(x,y,0)=\bw_{3}(x,y,h)=0, \text{ a.e. } (x,y)\in\mathbb{R}^2.
      \end{aligned}
      \right\}.
\]
and
\[ V=\left\{ \bu=\bw|_{\Omega};\;
      \begin{aligned}
      &  \bw\in (H^{1}_\loc(\RR^2\times(0,h)))^{3}, \;\bnabla\cdot\bw=0, \;\\
      &   \bw(x+L_{x},y,z)=\bw(x,y,z),\\
      &  \bw(x,y+L_{y},z)=\bw(x,y,z), \text{ a.e. } (x,y,z)\in \mathbb{R}^2\times (0,h).\\
      &  \bw(x,y,0)=\bw(x,y,h)=0, \text{ a.e. } (x,y)\in\mathbb{R}^2.
      \end{aligned}
      \right\}.
\]

The inner products in $H$ and $V$ are denoted respectively by
\[ (\bu,\bv)= \int_\Omega \bu(\bx)\cdot\bv(\bx) \;\rd\bx,
     \quad \Vinner{\bu,\bv} = \int_\Omega \sum_{i=1,3}
        \frac{\partial \bu}{\partial x_i}
          \cdot \frac{\partial \bv}{\partial x_i}\; \rd\bx,
\]
and the associated norms by $\left|\bu\right|_{0}=(\bu,\bu)^{1/2}$,
$\|\bu\|=\Vinner{\bu,\bu}^{1/2}$.

We identify $H$ with its dual and consider the dual space
$V'$ of $V$, so that $V\subseteq H \subseteq V'$, with the injections being
continuous, each space dense in the following one. We also denote by $H_\rw$
the space $H$ endowed with its weak topology.

We denote by $P_{\text{LH}}$ the (Leray-Helmhotz) orthogonal
projector in $L^2(\Omega)^3$ onto the subspace $H$. The operator
$A$ in \eqref{nseeq} is the Stokes operator given
by $A\bu = - P_{\text{LH}}\Delta \bu$. The term
$B(\bu,\bv) = P_{\text{LH}}((\bu\cdot\bnabla)\bv)$ is a bilinear term
associated with the inertial term.
Moreover, since the Stokes operator is a positive self-adjoint operator on $H$,
we  consider its powers $A^s$, $s\in\RR$, with domain $D(A^s)$. We have $V=D(A^{1/2})$ and its dual $V'=D(A^{-1/2})$.

The Stokes operator possesses a complete orthonormal basis of eigenvectors in
$H$, $\{\mathbf{w}_{j,l,k}\}_{j,l,k}$, of the form
\begin{equation}
\label{ev}
  \bw_{j,l,k}(x,y,z)
    =\exp\left(i\pi\left(\frac{jx}{L_x}+\frac{ly}{L_y}\right)\right)
    \hat \bw_{j,l,k}(z),
\end{equation}
where $(j,l,k)\in \ZZ\times \ZZ \times \NN$,
$A\mathbf{w}_{j,l,k}=\lambda_{j,l,k}\mathbf{w}_{j,l,k}$, and each
$\hat \bw_{j,l,k}(z)$ satisfies a one-dimensional eigenvalue problem,
with $0<\lambda_{j,l,k}\rightarrow\infty$, when $j,l,k\rightarrow\infty$.
We write the spectral expansion of $\bu$ in this basis as
\begin{equation}
  \bu(x,y,z) = \sum_{j,l,k} \hat u_{j,l,k} \bw_{j,l,k}(x,y,z).
\end{equation}
To each eigenvalue $\lambda_{j,l,k}$ we associate a wavenumber
$\kappa=\kappa_{j,l,k}=\lambda_{j,l,k}^{1/2}$. Since $\bu\in V$ vanishes
on the top and bottom walls, Poincar\'e inequality applies, yielding a bound on
$\left|\bu\right|_{0}$ in terms of $\left\|\bu\right\|$. In fact, we have
precisely
\begin{equation}
  \left|\bu\right|_{0}^2\leq \lambda_{1}^{-1}\left\|\bu\right\|^2,
\end{equation}
where $\lambda_{1}=\pi^2/h^2$ is the smallest positive eigenvalue of the
Stokes operator on this geometry. The smallest positive wavenumber is
$\kappa_{1}=\lambda_1^{1/2}=\pi/h$.

We define the component $\bu_{\kappa}$ of the vector field $\bu$, for a single
wavenumber $\kappa$, by
\[ \bu_{\kappa}=\sum_{\kappa_{j,l,k}=\kappa}\hat{u}_{j,l,k}\mathbf{w}_{j,l,k},
\]
and the component $\bu_{\kappa',\kappa''}$ with a range of wave
numbers $\left[\kappa',\kappa''\right)$ by
\[ \bu_{\kappa',\kappa''}=\sum_{\kappa'\leq\kappa<\kappa''}\bu_{\kappa}.
\]
We then write the Navier-Stokes equations projected on those
components in the form
\begin{equation}  
  \label{projNS} 
  \ddt{\bu_{\kappa',\kappa''}} + \nu
  A\bu_{\kappa',\kappa''} + B(\bu,\bu)_{\kappa',\kappa''} =
  \left(\mathbf{f}_{P}\right)_{\kappa',\kappa''},
\end{equation}
where
\begin{equation}
\label{forcecomp}
  \left(\mathbf{f}_{P}\right)_{\kappa}=\sum_{\kappa_{j,l,k}=\kappa}
    (\mathbf{f}_{P},\mathbf{w}_{j,l,k})\mathbf{w}_{j,l,k}.
\end{equation}

Now, taking the inner product in $H$ of the bilinear term with a
third variable yields a trilinear term
\[
 b(\bu,\bv,\bw) = (B(\bu,\bv),\bw),
\]
which is defined for $\bu,\bv,\bw$ in $V$. An important relation for
the trilinear term is the orthogonality property
\begin{equation}
b(\bu,\bv,\bv) = 0,
\end{equation}
for $\bu,\bv\in V$. It follows from this relation the anti-symmetry
property
\begin{equation}
\label{asym}
b(\bu,\bv,\bw) = - b(\bu,\bw,\bv),
\end{equation}
for $\bu,\bv,\bw\in V$.

\section{Statistical solutions and the Reynolds Equations}

A mathematical framework for the conventional theory of turbulence
is based on the concept of stationary statistical solution of
the Navier-Stokes equations. This amounts to considering the space $H$
as a probability space with the $\sigma$-algebra of the Borel
sets of $H$ and endowed with a Borel probability measure. The ensemble averages
are then regarded as averages with respect to this Borel probability measure.  In our
three-dimensional case we work mostly with the weak topology.
Fortunately the Borel $\sigma$-algebra generated by the weakly
open sets coincides with that for the open sets in the strong
topology.  Since $H$ is a separable Hilbert space every Borel
measure is automatically regular. An important consequence of the
regularity of a Borel probability measure is the density of the
continuous functions (or just weakly continuous functions) in the
space of integrable functions.

We say that a measure $\mu$ in $H$ is carried by a measurable set $E$
when $E$ has full measure in $H$, i.e.
$\mu\left( H \setminus E\right)=0$. The support of a Borel probability
measure $\mu$ is the smallest closed set which carries $\mu$. The ensemble
averages are regarded as averages with respect to a Borel
probability measure $\mu$ on $H$. If $\varphi:H\rightarrow \mathbb{R}$ is
a Borel function representing some physical information
$\varphi\left(\bu\right)$ extracted from a velocity field $\bu$,
such as kinetic energy, velocity, enstrophy, etc., then its mean
value is
\begin{equation}
\label{phiaverage}
\average{\varphi}=\int_{H}\varphi(\bu)d\mu(\bu).
\end{equation}
The reader is referred to \cite{fmrt} for more details.

Now, we define a class of Borel functions that are particularly
useful in order to make a rigorous definition of a stationary
statistical solution of the Navier-Stokes equations.
\begin{test}
We define the class $\mathcal{T}$ of test functions to be the set
of real-valued functionals $\Psi=\Psi(\bu)$ on $H$ that are
bounded on bounded subsets of $H$ and such that the following
conditions hold:
\begin{enumerate}
\item
For any $\bu\in V$, the Fr\'echet derivative $\Psi'(\bu)$ taken in
$H$ along $V$ exists. More precisely, for each $\bu\in V$, there
exists an element in $H$ denoted $\Psi'(\bu)$ such that
\begin{equation}
\frac{\left|\Psi(\bu+\bv)-\Psi(\bu)-(\Psi'(\bu),\bv)\right|_{0}}{\left|\bv\right|_{0}}\rightarrow
0\quad\text{as}\; \left|\bv\right|_{0}\rightarrow 0, \bv \in V.
\end{equation}
\item
$\Psi'(\bu)\in V$ for all $\bu\in V$, and $\bu\rightarrow
\Psi'(\bu)$ is continuous and bounded as a function from $V$ into
$V$.
\end{enumerate}
\end{test}

For example, we can take the cylindrical test functions
$\Psi:H\rightarrow\mathbb{R}$ of the form
$\Psi(\bu)=\psi\left((\bu,\mathbf{g}_{1}),\ldots,(\bu,\mathbf{g}_{m})\right)$,
where $\psi$ is a $C^{1}$ scalar function on $\mathbb{R}^{m}$,
$m\in\mathbb{N}$, with compact support, and
$\mathbf{g}_{1},\ldots,\mathbf{g}_{m}$ belong to $V$. For this
case we have
\[
\Psi'(\bu)=\sum_{j=1}^{m}\partial_{j}\psi((\bu,\mathbf{g}_{j}),\ldots,(\bu,\mathbf{g}_{j}))\mathbf{g}_{j},
\]
where $\partial_{j}\psi$ denotes the derivative of $\psi$ with
respect to the $j$-th variable. It follows that $\Psi '(\bu)\in V$
since it is a linear combination of the $\mathbf{g}_{j}$.

Now, we define the notion of a stationary statistical solution of
the Navier-Stokes equations.

\begin{sss}
\label{sss}
A stationary statistical solution of the Navier-Stokes equation is
a Borel probability measure $\mu$ on $H$ such that
\begin{enumerate}
\item
$\displaystyle{\int_{H}\left\|\bu\right\|^{2}d\mu(\bu)<\infty;}$
\item
$     \displaystyle{\int_{H}(\mathbf{F}(\bu),\Psi'(\bu))d\mu(\bu)=0,}
      \text{ for any } \Psi\in\mathcal{T},\; \text{where} 
      \;\mathbf{F}(\bu)\; \text{is as in} \;\eqref{nseeq}; $
 \item
$
\displaystyle{\int_{e_{1}\leq\left|\bu\right|_0^{2}/2<e_{2}}\left\{\nu\left\|\bu\right\|^{2}-(\mathbf{f}_{P},\bu)\right\}d\mu(\bu)\leq
  0}, \text{ for all }\; 0\leq e_1 < e_2\leq +\infty$.

\end{enumerate}

\end{sss}

The first condition means that an arbitrary stationary statistical solution has finite mean enstrophy. This is natural when we compare with individual solutions whose time average is bounded uniformly with respect to the time interval. It is
also needed to make sense out of the second condition.

The last condition on the definition above is an
energy-type inequality, and one can deduce from it that the support
of a stationary statistical solution is included in the weak
attractor $\mathcal{A}_{w}$, see \cite{fmrt, foias85},  which is bounded in $H$ according to
\begin{equation}
\label{attractorbound} \left|\bu\right|_{0}\leq\frac{\nu
G^{*}}{\kappa_1^{1/2}}=\frac{\nu h^{1/2}}{\pi^{1/2}}G^{*}, \quad \forall\bu\in
\mathcal{A}_{w},
\end{equation}
\[
G^{*}=\frac{h^{1/2}}{\nu^{2}\pi^{1/2}}|A^{-1/2}\bbf_P|_0,
\]
where $G^{*}$ is a nondimensional number called Grashof number.

The concept of stationary statistical solution is regarded as a
generalization of the notion of invariant measure. It is relevant
to our three-dimensional case, in which a semigroup is not
well-defined.

Due to these regularity properties obtained for stationary
statistical solutions (finite mean enstrophy and with support bounded
in $H$), the mean value $\average{\varphi(\bu)}$ can
be defined not only for weakly continuous functions bounded in $H$
but for any real-valued function $\varphi$ which is continuous in
$V$ and satisfies the estimate
\begin{equation}
\label{testcond}
\left|\varphi\right|_{0}\leq
C(\left|\bu\right|_{0})(1+\nu^{-2}\lambda_1^{1/2}\left\|\bu\right\|^{2}),
\end{equation}
where $C(\left|\bu\right|_{0})$ is bounded on bounded subsets of $H$.
Important examples of such $\varphi$ are $\left|\bu\right|_{0}^2$,
$\left\|\bu\right\|^2$, $
b(\bu_{\kappa_{0},\kappa},\bu_{\kappa_{0},\kappa},\bu_{\kappa,\infty})$,
and
$b(\bu_{\kappa,\infty},\bu_{\kappa,\infty},\bu_{\kappa_{0},\kappa})$.

By a duality argument we can extend the ensemble averages to functions with
value in some function space. More precisely, we define the velocity field
$\average{\bu}$ and the mean value $\average{B(\bu,\bu)}$ of the inertial
term by
\[ \left(\average{\bu},\bv\right)=\int_{H}(\bu,\bv)d\mu(\bu),\qquad\forall \bv\in V',
\]
\[ (\average{B(\bu,\bu)},\bv)=\int_{H}(B(\bu,\bu),\bv)d\mu(\bu), \qquad
       \forall \bv\in D\left(A^{3/8}\right).
\]
The mean flow $\average{\bu}$ is a vector field on $\Omega$ with
$\average{\bu}\in V$, while $\average{B(\bu,\bu)}\in D(A^{-3/8})$.

Since we assume statistical equilibrium, the stationary form of
the Reynolds equations can be recovered within this framework; see
also \cite{rosa}:

\begin{Reynoldsequation}
Given a stationary statistical solution in the sense of Definition
\ref{sss} the following functional form of the Reynolds equations hold in $V'$:
\begin{equation}
\label{Reynoldseq}
\nu A\average{\bu}+\average{B(\bu,\bu)}=\mathbf{f}_{P}.
\end{equation}
\end{Reynoldsequation}

\proof
Let $\psi$ be a $C^{1}$ real-valued function with compact
support on $\mathbb{R}$. For any $\bv\in V$ and any wavenumber
$\kappa$, the function
$\Phi(\bu)=\psi((\bu,\bv_{\kappa_{1},\kappa}))$ is a cylindrical
test function. Thus,
\[
\int_{H}\psi'((\bu,\bv_{\kappa_{1},\kappa}))\left\{(\mathbf{f}_{P},\bv_{\kappa_{1},\kappa})-\nu(A\bu,\bv_{\kappa_{1},\kappa})-b(\bu,\bu,\bv_{\kappa_{1},\kappa})\right\}d\mu(\bu)=0.
\]
Let $\psi '$ converge pointwise to $1$ while being uniformly
bounded, so that at the limit we find
\[
\int_{H}\left\{(\mathbf{f}_{P},\bv_{\kappa_{1},\kappa})-\nu(A\bu,\bv_{\kappa_{1},\kappa})-b(\bu,\bu,\bv_{\kappa_{1},\kappa})\right\}d\mu(\bu)=0.
\]
For each fixed $\bv\in V$, we may let $\kappa$ go to infinity to
find (since $\mu$ has finite enstrophy and with support bounded in
$H$)
\[
\int_{H}\left\{(\mathbf{f}_{P},\bv)-\nu(A\bu,\bv)-b(\bu,\bu,\bv)\right\}d\mu(\bu)=0.
\]
which gives us the result. \qed

\medskip
We end this section with a result concerning the Grashof number $G^{*}$,
which yields a bound in $H$ on the weak attractor
$\mathcal{A}_{w}$ in terms of P:

\begin{grashoflemma}
We have, more explicitly,
\begin{equation}
\label{Grashof}
G^{*}=\frac{\sqrt{3}L_{y}^{1/2}h^2}{6\pi^{1/2}\nu^2L_{x}^{1/2}}P.
\end{equation}

\end{grashoflemma}

\proof

 Since $\mathbf{f}_{P}=(P/L_{x})\mathbf{e_{1}}$, we have
that
\[
A^{-1}\mathbf{f}_{P}=(\frac{P}{2L_{x}}z(h-z),0,0).
\]
Hence,
\begin{equation}
\label{Aestimate}
\left|A^{-1/2}\mathbf{f}_{P}\right|_{0}^2=(\mathbf{f}_{P},A^{-1}\mathbf{f}_{P})=\int_{\Omega}\frac{P}{L_{x}}\frac{P}{2L_{x}}z(h-z)dx=\frac{L_{y}h^{3}}{12
L_{x}}P^{2}.
\end{equation}
Taking the square root of the equality above and substituting in
the definition of the Grashof number give us the result.
\qed
\rmk
The vector-field $A^{-1}\mathbf{f}_P$ is directly related to the plane Poiseuille flow.
In fact, the plane Poiseuille flow is precisely
$\bu=A^{-1}\mathbf{f}_P/\nu=(Pz(h-z)/2\nu L_x,0,0)$; see Section $7$.

\section{Characteristic dimensions and nondimensional numbers}

The macroscopic characteristic length is considered to be $h$ and the
macroscale characteristic wavenumber is $\kappa_{0}=1/h$. The total mass of the fluid in the channel is $\rho_{0}L_{x}L_{y}h$, where $\rho_{0}$ denotes the uniform mass density of the fluid. Then, for a given stationary statistical
solution $\mu$, the corresponding mean kinetic energy per unit
mass and the mean energy dissipation rate per unit time and unit
mass are given respectively by
\[ e=\frac{1}{2L_{x}L_{y}h}\average{\left|\bu\right|_{0}^2}, \quad 
   \epsilon=\frac{\nu}{L_{x}L_{y}h}\average{\left\|\bu\right\|^2}.
\]

The mean longitudinal velocity is defined by
\begin{equation}
\label{velocityflux}
  U=\frac{1}{L_yh}\int_{H}
      \left(\int_{0}^{h}\int_{0}^{L_y}u_{1}(x,y,z)dydz\right)d\mu(\bu).
\end{equation}
Note that this definition makes sense and the expression does not depend on $x$
due to the incompressibility and boundary conditions.

With this velocity scale we may define the following Reynolds number
\begin{equation}
  \label{reynoldsflux}
  Re=\frac{Uh}{\nu}.
\end{equation}
A dimensionless ratio of the applied pressure gradient to the square of
the flow velocity scale is provided by the skin friction coefficient given by
\begin{equation}
  \label{friction}
  C_{f}=\frac{Ph}{L_{x}U^2}.
\end{equation}

Thanks to condition $\left(3\right)$ of Definition \ref{sss} and to the 
divergence-free and boundary conditions, the mean longitudinal velocity 
and the mean energy dissipation rate are related by
\begin{equation}
\label{thmbUP}
   \epsilon\leq\frac{UP}{L_x}.
\end{equation}

Now, suppose that $\bu(\bx,t)$ is a weak solution of the
Navier-Stokes equations \eqref{nseeq1} with initial condition
$\bu_{0}(\bx)$. We define the finite-time average longitudinal velocity by
\begin{equation}
  U_{T}=\frac{1}{L_yh}\frac{1}{T}\int_{0}^{T}
   \left(\int_{0}^{h}\int_{0}^{L_y}u_{1}(x,y,z,t)dydz\right) dt.
\end{equation}
This expression is also well defined and independend of $x$
due to the incompressibility and boundary conditions.

In the sequel, we will also consider the similarly defined time-averaged 
dissipation rate
\[ \frac{\nu}{L_{x}L_{y}h}\average{\left\|\bu\right\|^2}_{T}
     =\frac{\nu}{L_{x}L_{y}h}\frac{1}{T}\int_{0}^{T}\left\|\bu(t)\right\|^2dt,
\]
and time-averaged kinetic energy
\[ \frac{1}{2L_{x}L_{y}h}\average{\left|\bu\right|_0^2}_{T}
     =\frac{1}{2L_{x}L_{y}h}\frac{1}{T}\int_0^T\left|\bu(t)\right|_0^2dt.
\]
Notice that even if finite-time
averages are bounded, their long-time limits may not exist.

\section{Time averages and stationary statistical solutions}

Since the usual limit of long-time averaged quantities may not exist, 
we aim to obtain eventual bounds for these time averaged quantities. 
In this section, we will establish, via generalized limits, a rigorous
relationship between certain limits of these quantities and the stationary 
statistical solutions, see \cite{fmrt}.

For example, suppose we are interested in estimating the
upper limit of the time averaged velocity  of a weak solution 
$\bu(\bx,t)$ of \eqref{nseeq1}
\begin{equation}
  \label{dissipaim}
  U=\frac{1}{L_yh}\limsup_{T\rightarrow\infty}\frac{1}{T}
    \int_{0}^{T}\left(\int_{0}^{L_{y}}\int_{0}^{h}u_{1}(x,y,z,t)dydz\right) dt.
\end{equation}
Since an upper bound for the mean longitudinal velocity  $U$, associated with 
an arbitrary stationary statistical solution $\mu$, is derived in 
\eqref{meanvelocity1}, namely

\[
U_\mu\leq\frac{\sqrt{3}h^{2}}{6\nu L_{x}}P,
\]
we may establish a direct relation between the time average longitudinal velocity  \eqref{dissipaim} and the mean longitudinal velocity  associated with a specific stationary statistical solution $\mu_0$, such as
 \begin{equation}
\label{taexample}
\limsup_{T\rightarrow\infty}\frac{1}{T}\int_{0}^{T}\left(\int_{0}^{L_{y}}\int_{0}^{h}u_{1}(x,y,z,t)dydz\right) dt
=\int_{H}\left(\int_{0}^{L_{y}}\int_{0}^{h}u_{1}(x,y,z)dydz\right)d\mu_0,
\end{equation}
 in such a way that we can give an upper bound to \eqref{dissipaim} using \eqref{meanvelocity1}, obtaining

 \begin{equation}
\label{taexamplec}
\frac{1}{L_yh}\limsup_{T\rightarrow\infty}\frac{1}{T}\int_{0}^{T}\left(\int_{0}^{L_{y}}\int_{0}^{h}u_{1}(x,y,z,t)dydz\right) dt=U_{\mu_0}\leq\frac{\sqrt{3}h^{2}}{6\nu L_{x}}P.
\end{equation}

This relation between long-time averages and stationary statistical solutions is realized via the notion of generalized limit, which is defined as follows

\begin{Glimit}
 A generalized limit is any linear functional, denoted $\Lim$, defined 
 on the space $\mathcal{B}([0,\infty))$
 of all bounded real-valued functions on $[0,\infty)$ and satisfying
 \begin{enumerate}
 \item
   $\Lim g(T)\geq 0, \;\forall g\in\mathcal{B}([0,\infty))$ with $g(s)\geq0$,
   $\forall s\geq 0;$
 \item
  $\Lim g(T)=\lim_{T\rightarrow\infty} g(T), \;\forall g\in\mathcal{B}([0,\infty))$ \\
  such that the classical limit, denoted $\lim_{T\rightarrow\infty}$, exists.
  \end{enumerate}
\end{Glimit}

\rmk \label{Berc}
It can be shown that given a particular $g_0\in\mathcal{B}([0,\infty))$ and
a sequence $t_j\rightarrow\infty$ for which $g_0$ converges to a number $l$,
there exists a generalized limit $\Lim$ satisfying $\Lim g_0=l$;
see \cite{bercovici, fmrt}.

 \begin{HB}
 \label{HB}
 Let $\varphi\in C(H_w)$. Suppose that for every stationary statistical solution $\mu$, the associated average of $\varphi$ satisfies $\average{\varphi}\leq C_1$, for some constant $C_1$. Then, given a weak solution $\bw(\bx,t)$ defined on $\left[0,\infty\right)$, we have
 \begin{equation}
 \label{HBproved}
 \limsup_{T\rightarrow\infty}\frac{1}{T}\int_0^T\varphi(\bw(t))dt\leq C_1.
 \end{equation}
  Similarly, if for some constant $C_2$ we have$\average{\varphi}\geq C_2$ for
  every stationary statistical solution, then
 \begin{equation}
 \label{HBnproved}
 \liminf_{T\rightarrow\infty}\frac{1}{T}\int_0^T\varphi(\bw(t))dt\geq C_2.
 \end{equation}
  \end{HB}
 \proof
 We will prove inequality \eqref{HBproved}. Inequality \eqref{HBnproved} follows by a similar argument.

  Let $\bw_0=\bw(0)$.  Consider the set
  $$K_{w}=\left\{\bv\in H;\,\left|\bv\right|^2\leq\left|\bw_{0}\right|_0^2
  +\left|\mathbf{f}_{P}\right|_0^2/\nu^2\lambda_1^2\right\},$$ endowed with
  the weak topology of $H$. $K_w$ is compact in $H_w$ and is such that
  $\bw(t)\in K_w$, for all $t\geq 0$; see \cite{constantinfoias, temam}.

 Let $\psi\in C(K_w)$. Since $K_w$ is compact, the function $t\mapsto \psi(\bw(t))$ is continuous and bounded. Thus,

\[ g_0(t)=\frac{1}{t}\int_0^t\psi(\bw(s))ds
\]
makes sense, and is continuous and bounded for $t\geq 0$. Therefore, 
its generalized limit is well defined, and by Remark \ref{Berc}, if we 
choose a subsequence $t_j\rightarrow\infty$ for which $g_0(t_j)$ 
converges to $\limsup_{t\rightarrow\infty}g_0(t)$, there exists a 
generalized limit $\Lim$ satisfying

\[ \Lim\frac{1}{T}\int_{0}^{T}\psi(\bw(t))dt
   =\limsup_{t\rightarrow\infty}\frac{1}{t}\int_{0}^{t}\psi(\bw(s))ds.
\]

Now, we relate this generalized time average with stationary
statistical solutions. Since the weak solution $t\mapsto \bw(t)$ belongs 
to the compact set $K_{w}$ in $H_{w}$, see \cite{fmrt}, and since
\[ \psi\mapsto\Lim\frac{1}{T}\int_0^T \psi(\bw(t))dt
\] 
is a positive linear functional
on $\mathcal{C}\left(K_{w}\right)$, we use the Kakutani-Riesz
representation theorem, see \cite{yosida}, and conclude that there exists a
measure $\mu_0$ on $H$ such that
\begin{equation}
\label{KRcor}
\Lim\frac{1}{T}\int_{0}^{T} \psi(\bw(\bx,t))dt=\int_{H}\psi(\bu)d\mu_0(\bu),
\end{equation}
for all $\psi\in C(K_w)$.
It is shown in \cite{fmrt} that $\mu_0$ defined above is a stationary statistical
solution.

Therefore, since $\varphi|_{K_w}\in \mathcal{C}\left(K_{w}\right)$, 
and $\mu(H\setminus K_w)=0$, for every stationary statistical solution, 
see \cite{fmrt}, and in particular for $\mu_0$, we conclude that
\[
\limsup_{t\rightarrow\infty}\frac{1}{t}\int_{0}^{t}\varphi(\bw(x,t))dt
=\int_{H}\varphi(\bu)d\mu_0(\bu).
\]
   Thus, since $\average{\varphi}\leq C_1$ for all stationary statistical solution $\mu$, and in particular $\mu_0$, we have
   \[
   \limsup_{T\rightarrow\infty}\frac{1}{T}\int_0^T\varphi(\bw(t))dt=\int_{H}\varphi(\bu)d\mu_0(\bu)\leq C_1.\qed
   \]

 Now, returning to the mean longitudinal velocity  example, since
 $$\bu\mapsto \int_{0}^{L_{y}}\int_{0}^{h}u_{1}(x,y,z)dydz
 $$
 belongs to $C(K_w)$, there exists a stationary statistical solution $\mu_0$ satisfying \eqref{taexample}, which together with \eqref{meanvelocity1}, yields the upper bound \eqref{taexamplec}.

\rmk
\label{ultrmk}
Proposition \ref{HB} shows that every estimate involving the average of a
continuous quantity on $C(K_{w})$ can be stated as a superior or inferior 
limit of its time average.

 This is true for the energy injection term, $(\mathbf{f}_P,\bu)$, and also 
for the mean longitudinal velocity $U$. However, we are also interested in 
estimating quantities involving $\left|\bu\right|_0$ and 
$\left\|\bu\right\|$, which are not weakly continuous. Fortunately, we are 
still able to estimate these quantities by approximating via Galerkin 
projections as shown in the next proposition.

\begin{galerkapprox}
  \label{galerkapprox}
  Let $\bw(\bx,t)$ be a weak solution of the NSE defined on 
  $\left[0,\infty\right)$, and suppose that for every stationary statistical 
  solution $\mu$, we have the following bounds
  \[ C_1\leq\int_{H}\left|\bu\right|_0^2d\mu_0(\bu)\leq C_2,
  \]
  and
  \[ C_3\leq\int_{H}\left\|\bu\right\|^2d\mu_0(\bu).
  \]
  Then, we also have
 \begin{equation}
 \label{HBnprovedgal}
 \liminf_{T\rightarrow\infty}\frac{1}{T}\int_0^T\left|\bw(t)\right|_0^2dt\geq C_1,
 \end{equation}
    \begin{equation}
   \label{HBprovedgal}
 \limsup_{T\rightarrow\infty}\frac{1}{T}\int_0^T\left|\bw(t)\right|_0^2dt\leq C_2,
 \end{equation}
  and
\begin{equation}
 \label{epsHBnprovedgal}
 \liminf_{T\rightarrow\infty}\frac{1}{T}\int_0^T\left\|\bw(t)\right\|^2dt\geq C_3.
 \end{equation}
  \end{galerkapprox}
\proof

 Since $\left|P_{\kappa}\bu\right|_0, \left\|P_{\kappa}\bu\right\|\in C(K_w)$, where $P_\kappa$ are the usual Galerkin projectors, we have by \eqref{KRcor} that given a stationary statistical solution $\mu_0$ generated by a generalized time average, the following equations are valid
 \begin{equation}
\label{KRcoren}
\Lim\frac{1}{T}\int_{0}^{T} \left|P_\kappa\bw(t)\right|_0^2dt=\int_{H}\left|P_\kappa\bu\right|_0^2d\mu_0(\bu),
\end{equation}
 and
 \begin{equation}
\label{KRcorenst}
\Lim\frac{1}{T}\int_{0}^{T} \left\|P_\kappa\bw(t)\right\|^2dt=\int_{H}\left\|P_\kappa\bu\right\|^2d\mu_0(\bu).
\end{equation}
 Now, since
  $$
  \left|\bw(t)\right|_0^2-\left|P_\kappa\bw(t)\right|_0^2 =\left|Q_\kappa\bw(t)\right|_0^2 \leq \kappa^{-2}\|\bw(t)\|^2,
  $$
  and
$$
\frac{1}{T}\int_0^T\|\bw(t)\|^2dt\leq C<\infty,
$$
where $C$ is independent of $T$ \cite{constantinfoias, temam}, we have 
by the usual properties of the generalized limits that
 \begin{multline}
  \Lim\frac{1}{T}\int_{0}^{T} \left|\bw(t)\right|^2dt
     -\Lim\frac{1}{T}\int_{0}^{T} \left|P_\kappa\bw(t)\right|^2dt
   \leq\kappa^{-2}\Lim\frac{1}{T}\int_0^T\|\bw(t)\|^2dt \\
   \leq\kappa^{-2}\limsup\frac{1}{T}\int_0^T\|\bw(t)\|^2dt
   \leq C\kappa^{-2}\rightarrow 0,\quad  \kappa\rightarrow\infty.
\end{multline}
Thus, considering the generalized limit, $\Lim$, that extends the left
hand side of \eqref{HBnprovedgal}, we have
\begin{multline}
  \liminf_{T\rightarrow\infty}\frac{1}{T}\int_0^T\left|\bw(t)\right|_0^2dt
    =\Lim\frac{1}{T}\int_0^T\left|\bw(t)\right|_0^2dt
    =\lim_{\kappa\rightarrow\infty}\Lim\frac{1}{T}\int_{0}^{T} 
       \left|P_\kappa\bw(t)\right|^2dt \\
    =\lim_{\kappa\rightarrow\infty}\int_{H}\left|P_\kappa\bu\right|^2
        d\mu_0(\bu)
    =\int_{H}\left|\bu\right|^2d\mu_0(\bu)\geq C_1,
\end{multline}
where the last equality of the expression above follows from the 
Monotone Convergence Theorem. The bound \eqref{HBprovedgal} follows in a 
similar way.

Now, we will prove \eqref{epsHBnprovedgal}. Consider the generalized limit, $\Lim$, that extends the l.h.s. of \eqref{epsHBnprovedgal}, and notice that
\[
\left\|\bw(t)\right\| \geq \left\|P_\kappa \bw(t)\right\|.
\]
Then, we have
\begin{multline}
\liminf_{T\rightarrow\infty}\frac{1}{T}\int_0^T\left\|\bw(t)\right\|_0^2dt=\Lim\frac{1}{T}\int_0^T\left\|\bw(t)\right\|_0^2dt\geq\lim_{\kappa\rightarrow\infty}\Lim\frac{1}{T}\int_{0}^{T} \left\|P_\kappa\bw(t)\right\|^2dt \\
=\lim_{\kappa\rightarrow\infty}\int_{H}\left\|P_\kappa\bu\right\|^2d\mu_0(\bu)=\int_{H}\left\|\bu\right\|^2d\mu_0(\bu)\geq C_3,
\end{multline}
where, again, the last equality in the expression above follows from 
the Monotone Convergence Theorem.\qed

\rmk Propositions \ref{HB} and \ref{galerkapprox} show that, except for 
Theorem \ref{epsilonupperG} and Propositon \ref{corCD2}, every estimate 
in the sequel can be stated as a superior or inferior limit of their time 
averages. The reason why these results do not apply to Theorem 
\ref{epsilonupperG} and Propositon \ref{corCD2} is that they involve 
an upper bound to $\left\|u\right\|$, which is not considered by
the propositions above. However, they can still be stated in terms of 
their time averages as seen in Remark \ref{rmklab}.

\section{Estimates on the mean longitudinal velocity  and on the skin friction
  coefficient}

We start by deriving an upper bound on the mean longitudinal velocity.
\begin{thm1}
\label{meanvelocity1}
For every stationary statistical solution, the mean longitudinal velocity $U$ satisfies
\begin{equation}
 U\leq \frac{\sqrt{3}h^2}{6\pi\nu L_x}P.
\end{equation}
\end{thm1}

\proof
It follows directly from the definition of $U$ and from the Cauchy-Schwarz
and H\"older inequalities that
\begin{equation}
  \label{thm1aux1}
  U\leq\frac{1}{L_x^{1/2}L_y^{1/2}h^{1/2}}
    \average{\left|\bu\right|^{2}_0}^{1/2}.
\end{equation}
Now, by \eqref{attractorbound} and \eqref{Grashof}, we can estimate the 
term $\average{\left|\bu\right|^{2}_0}$ as follows
\begin{equation}
  \label{thm1aux2}
  \average{\left|\bu\right|_{0}^{2}}\leq \frac{\nu^{2}h}{\pi}{G^{*}}^{2}
   =\frac{L_yh^5}{12\pi^2\nu^{2}L_x}P^{2}.
\end{equation}
Substituting \eqref{thm1aux2} into \eqref{thm1aux1}, we obtain the result.\qed

A lower bound for the skin friction coefficient, $C_{f}$, follows directly from the
theorem above

\begin{coefupper}
For every stationary statistical solution, the skin friction coefficient, $C_{f}$, satisfies
\begin{equation}
\label{frictionestimate}
 C_{f}\geq \frac{2\pi\sqrt{3}}{Re}.
\end{equation}
\end{coefupper}
\proof
It follows immediately from Theorem \ref{meanvelocity1} that
\[ C_f=\frac{Ph}{L_xU^2}\geq\frac{Ph}{L_xU}\frac{6\pi\nu L_x}{\sqrt{3}h^2P}
   =2\pi\sqrt{3}\frac{\nu}{hU},
\]
and the result follows from the definition of the Reynolds number \eqref{reynoldsflux}. \qed

Now, we give a lower bound estimate for the mean longitudinal velocity $U$ 
following the calculations of \cite{constantindoering95}, but avoiding using 
an equation for the fluctuation $\bv$.

\begin{CDestimate}
\label{CDestimate}
For every stationary statistical solution, the mean longitudinal velocity $U$ 
satisfies
\[  U\geq \sup\left\{\frac{h^2}{12\nu
   L_{x}}P -\frac{\nu L_{x}}{Ph}\int_{0}^{h}\left(U_1'(z)-\frac{P}{2\nu
      L_{x}}\left(h-2z\right)\right)^2dz;\; \bU \in \Ucal \right\},
\]
where
\[ \Ucal = \left\{ \bU\in V;\; \bU(x,y,z)=(U_1(z),0,0), 
      \;U_1\in H_0^1(0,h), \;\average{H_{\bU}(\bu-\bU)}\geq 0 \right\},
\]
and
\[ H_{\bU}(\bu-\bU)
   =\nu\frac{\left\|\bu-\bU\right\|^2}{2}+b(\bu-\bU,\bU,\bu-\bU).
\]
\end{CDestimate}

\proof Let $\bU\in V$ be of the form $\bU(x,y,z)=(U_1(z),0,0)$. We have
\begin{equation}
  \label{CDstepp}
  \average{\left\|\bu-\bU\right\|^2}=\average{\left\|\bu\right\|^2}
    -2\average{\Vinner{\bu,\bU}}+\left\|\bU\right\|^2.
\end{equation}
Now, since $\bU\in V$ is fixed, we can multiply it with the Reynolds 
equations and obtain
\begin{equation}
\label{CDstep2}
\nu\average{\Vinner{\bu,\bU}}=(\mathbf{f}_{P},\bU)-\average{b(\bu,\bu,\bU)}.
\end{equation}

Substituting \eqref{CDstep2} into \eqref{CDstepp}, we obtain

\begin{equation}
\nu\average{\left\|\bu\right\|^2}=\nu\average{\left\|\bu-\bU\right\|^2}+2\left((\mathbf{f}_{P},\bU)-\average{b(\bu,\bu,\bU)}\right)-\nu\average{\left\|\bU\right\|^2}.
\end{equation}

Since $\bU(x,y,z)=(U_1(z),0,0)$,
by the anti-symmetry property of the trilinear term and by \eqref{thmbUP}, 
we have

\begin{equation}
\label{barudissiplow}
\begin{aligned}
  L_{x}L_{y}hU&\geq\frac{\nu L_x}{P}\average{\left\|\bu\right\|^2}\\
&   =\frac{2L_x}{P}\average{\nu\frac{\left\|\bu-\bU\right\|^2}{2}+b(\bu,\bU,\bu)}+2\frac{L_x}{P}(\mathbf{f}_{P},\bU)-\frac{\nu L_x}{P}\left\|\bU\right\|^2.
\end{aligned}
\end{equation}

Due to the form of $\bU$, we have $b(\bU,\bU,\bu)=0$. 
The orthogonality property implies 
\[b(\bu-\bU,\bU,\bU)=0.
\]
Thus,
\[ b(\bu,\bU,\bu)=b(\bu-\bU,\bU,\bu-\bU),\quad \forall \bu\in V.
\]

Since $\mu$ is carried by $V$, we find

\begin{multline}
 \left(L_{x}L_{y}h\right)U \\
   =\frac{2L_x}{P}\average{\nu\frac{\left\|\bu-\bU\right\|^2}{2}
     +b(\bu-\bU,\bU,\bu-\bU)}
     +\frac{2L_x}{P}(\mathbf{f}_{P},\bU)
     -\frac{\nu L_x}{P}\left\|\bU\right\|^2.
\end{multline}

Thus, considering only background flows $\bU$ 
such that $\average{H_{\bU}(\bv)}\geq 0$, we have
\begin{equation}
\label{epslow}
  \left(L_{x}L_{y}h\right)U\geq \frac{2L_x}{P}(\mathbf{f}_{P},\bU)
   -\frac{\nu L_{x}}{P}\left\|\bU\right\|^2.
\end{equation}
We obtain the result by completing the squares.

\rmk This theorem was shown in \cite{constantindoering95}, in the context 
of long time averages. It was obtained from a derivation of an energy equation
for the fluctuation variable $\bv=\bu-\bU$:
\begin{equation}
\label{CDfluct}
\frac{1}{2}\frac{d}{dt}\left|\bv\right|_0^2+\nu\left\|\bv\right\|^2+\nu\Vinner{\bv,\bU}+b(\bU,\bU,\bv)+b(\bv,\bU,\bv)=(\mathbf{f}_{P},\bv)
  \end{equation}
 and by considering an energy equation for $\bu$:
\begin{equation}
\label{CDhyp}
\frac{d}{dt}\frac{1}{2}\left|\bu\right|_0^2+\nu\left\|\bu\right\|^2=(L_{y}h)PU.
\end{equation}
Taking the long time average in both sides of \eqref{CDhyp}, considering the same hypothesis for $\bU$, and
substituting it into \eqref{CDfluct}, they have obtained the
corresponding result for long time averages.

However, since we want to consider any
stationary statistical solutions and general
weak solutions of the Navier-Stokes equations, we treat carefully 
the fluctuation component and avoid the energy equation \eqref{CDfluct}.  
The slightly modified and simpler proof presented in Theorem
\ref{CDestimate} achieves this aim.

\begin{corCD}
\label{corCD}
For every stationary statistical solution, the mean longitudinal velocity 
and the skin friction coefficient satisfy
\begin{equation} U
     \geq \begin{cases}
          \label{CDphilow}\displaystyle \frac{2}{27}\frac{h^2}{\nu L_x}P,
          & \displaystyle
              \text{if }\; 0<P\leq \frac{27\sqrt{2}\pi^2\nu^2L_x}{4h^3};
                \medskip \\
          \displaystyle
             \frac{2^{5/4}\pi}{3^{3/2}}\frac{h^{1/2}}{L_x^{1/2}}P^{1/2}
                 - \frac{\sqrt{2}\pi^2}{2}\frac{\nu}{h} ,
          & \displaystyle
              \text{if }\; P> \frac{27\sqrt{2}\pi^2\nu^2L_x}{4h^3}.
       \end{cases}
\end{equation}

and

\begin{equation} C_f \leq \begin{cases}
    \label{CDcflow}\displaystyle \frac{27}{2}\frac{1}{\Rey},
      & \displaystyle \text{if }\; 0<P\leq
           \frac{27\sqrt{2}\pi^2\nu^2L_x}{4h^3}; \medskip \\
      \displaystyle \frac{27\sqrt{2}}{8\pi^2}
        \left(1 + \frac{\sqrt{2}\pi^2}{2}\frac{1}{\Rey} \right)^2,
     & \displaystyle
         \text{if }\; P\geq \frac{27\sqrt{2}\pi^2\nu^2L_x}{4h^3}.
    \end{cases}
\end{equation}
\end{corCD}

\proof Following Constantin and Doering, consider the background-flow 
of the form $\bU(x,y,z)=(U_1(z),0,0)$ with 
\[ U_1(z)= \begin{cases}
    \displaystyle \frac{V}{\delta}z, & 0\leq z \leq \delta; \\
     V, & \delta\leq z \leq h-\delta; \\
    \displaystyle \frac{V}{\delta}\left(h-z\right), & h-\delta\leq z \leq h.
  \end{cases}
\]

We will verify that this flow satisfies the spectral constraint
$H_{\bU}(\bv)\geq 0$, for appropriate choices of $V$ and $\delta$. 
For that purpose, we bound the integral of
$U_1'(z)v_{1}v_{3}$ in terms of $\delta$ and $\left\|\bv\right\|^2$.
First, divide this integral into two parts, one from $0$ to $\delta$, and
the other from $h-\delta$ to $h$.

In order to bound the first integral,
    consider the spaces $\tilde{H}=L^2(0,\delta)$, with the usual
    $L^2$ inner product, and $\tilde{V}=\left\{\bu\in H^1(0,\delta);
    u(0)=0 \right\}$, with the inner product
    $\Vinner{u,v}=\int_{0}^{\delta}u'(z)v'(z)dz$.
Consider the operator $\tilde{A}:\tilde{V}\rightarrow \tilde{H}$ defined by
    \[ (\tilde{A}u,v)=\Vinner{u,v},\quad \forall v\in \tilde{V},
    \]
    and 
    $D(\tilde{A})=\left\{u\in \tilde{V};\; \tilde{A}u\in \tilde{H}\right\}$. 
    One can
    show that $\tilde{A}$  is self-adjoint and invertible, with compact
    inverse, and that the smallest associated eigenvalue is
    $\tilde\lambda_{1}=\pi^2/4\delta^2$. Therefore,

\begin{equation}
  \tilde\lambda_{1}\int_{0}^{\delta}\left|u(z)\right|^2dz\leq\int_{0}^{\delta}
     \left|\frac{\partial u(z)}{\partial z}\right|^2dz.
\end{equation}

A similar statement can be made for the integral between $h-\delta$ and
$h$. Thus, the integral of $U'(z)v_{1}v_{3}$ can be estimated in the
following way

$\displaystyle{\left|\int_{0}^{L_{x}}\int_{0}^{L_{y}}\int_{0}^{h}U'(z)v_{1}v_{3}dxdydz\right|}$

\begin{eqnarray*}
\phantom{\leq\leq}&\leq &\frac{V}{\delta}\left|\int_{0}^{L_{x}}\int_{0}^{L_{y}}\int_{0}^{\delta}v_{1}v_{3}dxdydz-\int_{0}^{L_{x}}\int_{0}^{L_{y}}\int_{h-\delta}^{h}v_{1}v_{3}dxdydz
         \right| \\
\phantom{\leq\leq} & \leq & \frac{V}{\delta}\int_{0}^{L_{x}}\int_{0}^{L_{y}}\int_{0}^{\delta}
      \alpha\frac{\left|v_{1}\right|^2}{2}+\frac{\left|v_{3}\right|^2}{2\alpha}dxdydz\\
    \phantom{\leq\leq} & \phantom{\leq} &  \qquad + \frac{V}{\delta}\int_{0}^{L_{x}}\int_{0}^{L_{y}}\int_{h-\delta}^{h}
      \alpha\frac{\left|v_{1}\right|^2}{2}+\frac{\left|v_{3}\right|^2}{2\alpha}dxdydz
 \\
   \phantom{\leq\leq} & \leq &
   \frac{2V\delta}{\pi^2}\left(\alpha\left|\frac{\partial
         v_{1}}{\partial
         z}\right|_{0}^2+\frac{1}{\alpha}\left|\frac{\partial
         v_{3}}{\partial z}\right|_{0}^2 \right)\\
 \phantom{\leq\leq} & \leq &
   \frac{2V\delta}{\pi^2}\left(\alpha\left|\frac{\partial
         v_{1}}{\partial z}\right|_{0}^2+
         \frac{1}{2\alpha}\left(\left|\frac{\partial v_{3}}{\partial z}\right|_{0}^2
      + \left|\frac{\partial v_{1}}{\partial x}\right|_{0}^2
      + \left|\frac{\partial v_{2}}{\partial y}\right|_{0}^2
      + \left|\frac{\partial v_{1}}{\partial y}\right|_{0}^2
      + \left|\frac{\partial v_{2}}{\partial x}\right|_{0}^2 \right)\right)\\
   \end{eqnarray*}

   The last step above follows from the following inequality stated in \cite{constantindoering94}:

   \[
   \left|\frac{\partial v_{3}}{\partial z}\right|_{0}^2
   \leq
   \frac{1}{2}\left(\left|\frac{\partial v_{3}}{\partial z}\right|_{0}^2
      + \left|\frac{\partial v_{1}}{\partial x}\right|_{0}^2
      + \left|\frac{\partial v_{2}}{\partial y}\right|_{0}^2
      + \left|\frac{\partial v_{1}}{\partial y}\right|_{0}^2
      + \left|\frac{\partial v_{2}}{\partial x}\right|_{0}^2
      \right),
   \]
   which is valid for divergence-free vector fields.

Thus, choosing $\alpha=\sqrt{2}/2$, we have

\[
\left|\int_{0}^{L_{x}}\int_{0}^{L_{y}}\int_{0}^{h}U_1'(z)v_{1}v_{3}dxdydz\right|\leq\frac{\sqrt{2}}{\pi^2}V\delta\left\|\bv\right\|^2.
\]

Hence, $H_{\bU}(\bv)$ is bounded from below by

\[
H_{\bU}(\bv)\geq\left(\frac{\nu}{2}-\frac{\sqrt{2}}{\pi^2}V\delta\right)\left\|\bv\right\|^2.
\]

Therefore, $H_{\bU}$ is non-negative if $\delta\leq\nu\pi^2/2\sqrt{2}V$, with $V$ sufficiently large to fulfill the compatibility hypothesis $\delta<h/2$.
Now, by substituting $\bU$ in $\eqref{epslow}$, we give a lower bound for
$U$:
\begin{equation}
\label{barulower}
  U\geq 2L_xL_y\left(hV-\delta V-\frac{\nu L_xV^2}{\delta P}\right).
\end{equation}
We maximize the lower bound above, respecting the compatibility hypotheses, with the following choices of $V$ and $\delta$:
\begin{equation}
\label{baruloweraux1}
  V=\frac{\pi h^{1/2}}{3^{1/2}2^{3/4}L_x^{1/2}}P^{1/2},\quad
  \delta=\frac{3^{1/2}\nu L_x^{1/2}}{2^{3/4}h^{1/2}P^{1/2}}
    \quad\quad\text{if}\quad P>27\frac{\sqrt{2}\pi^2\nu^2L_x}{4h^3},
\end{equation}
and
\begin{equation}
\label{baruloweraux2}
V=\frac{h^2P}{9\nu L_x},\quad\delta=\frac{h}{3}\quad\quad \text{if}\quad P\leq27\frac{\sqrt{2}\pi^2\nu^2L_x}{4h^3}.
\end{equation}
The result follows immediately from the substitution of \eqref{baruloweraux1} and \eqref{baruloweraux2} into \eqref{barulower}.\qed

\rmk
  Theorem \ref{corCD} gives a uniform upper bound estimate
  for the skin friction coefficient for high Reynolds numbers.
  Even though this constant upper
  bound estimate is predicted by the Kolmogorov theory of homogeneous
  turbulence, it is known from experiments that
  corrections are necessary for turbulence in the presence of walls, see
  \cite{constantindoering94, constantindoering95, Prandtl, TL}.
  Actually, closure approximation theories establish the following logarithmic
  friction law which has been confirmed by high Reynolds number pipe flow
  experiments:
  \begin{equation}
  C_{f}\sim \frac{1}{(\ln{R})^2}.
  \end{equation}
  Thus, we conclude that while empirical arguments and experimental
  data predicts a logarithmic friction law, our rigorous mathematical
  bounds can only assert that
  \begin{equation}
    \frac{2\pi\sqrt{3}}{\Rey}
   \leq C_{f} \leq \frac{27\sqrt{2}}{8\pi^2} + O\left(\frac{1}{\Rey^2}\right).
  \end{equation}
  The lower bound for $C_f$ is of the order of the skin friction 
  coefficient for the plane Poiseuille flow; see Section \ref{poiseuillesec}.

\rmk Note also that for high Reynolds number flows, the characteristic
  background velocity which leads to the estimate above is of the order of
  \[ V \sim (h/L_x)^{1/2}P^{1/2},
  \]
  while the corresponding ``boundary layer'' 
  length is of the order of
  \[ \delta \sim (\nu L_x^{1/2}/h^{1/2})P^{-1/2}.
  \]

\section{Other Estimates}

We start by deriving a lower bound for the energy dissipation rate
$\epsilon$.

\begin{thm1}
\label{thm1}
For every stationary statistical solution, the energy dissipation
rate satisfies
\begin{equation}
\label{epsilonL}
 \epsilon
     \geq \begin{cases}
          \displaystyle \frac{2}{27}\frac{h^2}{\nu L_x^2}P^2,
          & \displaystyle
              \text{if }\; 0<P\leq \frac{27\sqrt{2}\pi^2\nu^2L_x}{4h^3},
                \medskip \\
          \displaystyle
             \frac{2^{5/4}\pi}{3^{3/2}}\frac{h^{1/2}}{L_x^{3/2}}P^{3/2}
                 - \frac{\sqrt{2}\pi^2}{2}\frac{\nu}{h} ,
          & \displaystyle
              \text{if }\; P> \frac{27\sqrt{2}\pi^2\nu^2L_x}{4h^3}.
       \end{cases}
\end{equation}
\proof
The result follows from noticing that the estimate \eqref{barudissiplow} obtained in Proposition \ref{CDestimate} is actually a lower bound for $\average{\left\|\bu\right\|^2}$, and, therefore, we can follow the subsequent calculations in the exact same way with this term instead of $U$.\qed
\end{thm1}

Now, we give a lower bound on the mean kinetic energy.

\begin{cor2}
For every stationary statistical solution, the mean kinetic energy
$e$ satisfies
\begin{equation}
\label{energyL}
e\geq \frac{h}{6L_{x}}P-\frac{4\nu}{L_{y}h^3}\left(\frac{\nu^{2}L_{y}^{2}h^{2}}{P}+\frac{L_{y}^{2}h^{5}}{12 L_{x}}\right)^{1/2}P^{1/2}+\frac{4\nu^{2}}{h^{2}}.
\end{equation}
\end{cor2}

\proof

Taking the inner product with $A^{-1}\mathbf{f}_{P}$ in the Reynolds equation
yields
\begin{equation}
\label{nonlineq}
       \left|A^{-1/2}\mathbf{f}_{P}\right|_{0}^2
      =\nu\average{\left(\bu,\mathbf{f}_{P}\right)}+b\left(\bu,\bu,A^{-1}\mathbf{f}_{P}\right)\leq
       \nu\average{\left|\bu\right|_{0}}\left|\mathbf{f}_{P}\right|_{0}+\average{\left|b\left(\bu,A^{-1}\mathbf{f}_{P},\bu\right)\right|},
       \end{equation}
and since

\begin{equation}
\label{epsiloncalculations}
\begin{aligned}
       \average{\left|b\left(\bu,A^{-1}\mathbf{f}_{P},\bu\right)\right|}
    &  =\average{\left|\int_{\Omega}u_{3}\left(\frac{\partial}{\partial z}\frac{P}{2L_{x}}z(h-z)\right)u_{1}dx\right|}\\
    &  \leq\frac{P}{2L_{x}}\average{\int_{\Omega}\left|h-2z\right|\left|u_{3}\right|\left|u_{1}\right|dx}\\
    &  \leq\frac{Ph}{4L_{x}}\average{\int_{\Omega}\left|u_{3}\right|^{2}+\left|u_{1}\right|^{2}dx}\\
    &
    \leq\frac{Ph}{4L_{x}}\average{\left|\bu\right|_{0}^{2}},
\end{aligned}
\end{equation}
we find from \eqref{nonlineq} that
\begin{multline}
      \left|A^{-1/2}\mathbf{f}_{P}\right| _{0}^{2} \leq\nu\left|\mathbf{f}_{P}\right|\average{\left|\bu\right|_{0}^{2}}^{1/2}+\average{\left|b\left(\bu,A^{-1}\mathbf{f}_{P},\bu\right)\right|_{0}}\\
     \leq    \nu \frac{PL_{y}^{1/2}h^{1/2}}{L_x^{1/2}}\average{\left|\bu\right|_{0}^{2}}^{1/2} +P \frac{h}{4L_{x}}\average{\left|\bu\right|_{0}^{2}}
\end{multline}

Then, using \eqref{Aestimate}, we have
\begin{equation}
\label{Pestimate}
\frac{L_{y}h^{3}}{12L_{x}}P \leq     \frac{\nu}{L_{x}}\average{\left|\bu\right|_{0}^{2}}^{1/2} +
\frac{h}{4L_x}\average{\left|\bu\right|_{0}^{2}},
\end{equation}
which is of the form $ar^{2}+br+c\geq 0$ for $r=\average{\left\|\bu\right\|^{2}}^{1/2}$,
$
a=h/4L_x,\;  b=\nu/L_x,\; c=(L_yh^3)/12L_x .
$
It gives us $ r^{2}\geq b^{2}/2a^{2}-b\left(b^{2}+4ac\right)^{1/2}/a+c/a$, which implies

\[ \average{\left|\bu\right|_{0}^{2}}\geq \frac{h^2L_y}{3}P-\frac{8\nu L_x}{h^2}\left(\frac{\nu^{2}L_{y}^{2}h^{2}}{P}+\frac{L_{y}^{2}h^{5}}{12
      L_{x}}\right)^{1/2}P^{1/2}+\frac{8\nu^{2}L_x L_y}{h}.
\]
 \qed

\begin{enupper}
\label{enupper}
For every stationary statistical solution, the mean kinetic energy
$e$ satisfies
\begin{equation}
   e\leq \frac{h^4}{24\pi^2\nu^2 L_x^2}P^2.
\end{equation}
\end{enupper}
\proof This follows directly from inequality \eqref{thm1aux2}.\qed

\begin{thm2}
For every stationary statistical solution, the energy dissipation
rate $\epsilon$ satisfies

\begin{equation}
  \label{epsilonupperG} 
  \epsilon\leq \frac{\sqrt{3}h^{2}}{6\pi \nu L_x^2}P^2.
\end{equation}

\end{thm2}

\proof

This follows directly from inequalities \eqref{thmbUP} and \eqref{meanvelocity1}.\qed

Now, we state a partial
rigorous confirmation of the Kolmogorov dissipation law in terms of $U$.

\begin{corCD2}
For every stationary statistical solution, and sufficiently large pressure drop P, namely $P\geq 27\sqrt{2}\pi^2\nu^2L_x/4h^3$, the associated energy
dissipation rate satisfies
\begin{equation}
\label{corCD2}
\epsilon\leq \left(\frac{3}{2^{5/2}\pi^2}+\frac{27\pi}{4}\frac{1}{\Rey}+\frac{27\pi^2}{2^{7/2}}\frac{1}{\Rey^2}\right)\frac{U^3}{h}.
\end{equation}
\end{corCD2}

\proof
Taking the square of both sides of the second inequality in \eqref{CDphilow}, we have
\begin{equation}
\label{plowerbaru}
P\leq \frac{3L_x}{2^{5/2}\pi^2h}U^2 + \frac{27L_x\pi\nu}{4h^2}U+\frac{27\pi^2L_x\nu^2}{2^{7/2}h^3}.
\end{equation}
 Substituting \eqref{plowerbaru} into \eqref{thmbUP}, we obtain the result.\qed

\smallskip

\rmk
\label{rmklab}
Note that we cannot invoke Proposition \ref{HB} neither Proposition \ref{galerkapprox} to state the results \eqref{corCD2} and \eqref{epsilonupperG} in terms of their time averages. However, we can improve these results as follows.

Let $\bu(\bx,t)$ be a weak solution of the Navier-Stokes equations. It follows from the classical energy inequality for weak solutions of the NSE defined on
$\left[0,\infty\right)$, see \cite{constantinfoias, temam}, that
\begin{equation}
\limsup_{T\rightarrow\infty}\frac{1}{T}\int_{0}^{T}\nu\left\|\bu(s)\right\|^2ds\leq\liminf_{T\rightarrow\infty}\frac{1}{T}\int_{0}^{T}\left(\mathbf{f}_{P},\bu(s)\right)ds.
\end{equation}
Then, an inequality similar to \eqref{thmbUP} can be stated
\begin{equation}
\label{evbdext}
\frac{\nu}{L_xL_yh}\limsup_{T\rightarrow\infty}\average{\left\|\bu(s)\right\|^2}_T\leq\frac{P}{L_{x}}\liminf_{T\rightarrow\infty}U_T.
\end{equation}

Hence, if we consider inequality \eqref{meanvelocity1} for the
stationary statistical solution $\mu_0$ that extends the
inferior limit of the time-averaged longitudinal velocity of $\bu(\bx,t)$, we have
\[
\frac{\nu}{L_xL_yh}\limsup_{T\rightarrow\infty}\average{\left\|\bu(s)\right\|^2}_T\leq\frac{P}{L_{x}}\liminf_{T\rightarrow\infty}U_T=\frac{PU_{\mu_0}}{L_x}\leq\frac{\sqrt{3}h^{2}}{6\nu L_{x}^{2}}P^2.
\]

Similarly, considering inequality \eqref{CDphilow} for the same $\mu_0$ above, we have

\begin{equation}
\label{evbdext2}
\begin{aligned}
P &\leq \frac{3L_x}{2^{5/2}\pi^2h}U_{\mu_0}^2 + \frac{27L_x\pi\nu}{4h^2}U_{\mu_0}+\frac{27\pi^2L_x\nu^2}{2^{7/2}h^3}\\
\phantom{P}&= \frac{3L_x}{2^{5/2}\pi^2h}(\liminf_{T\rightarrow\infty}U_T)^2 + \frac{27L_x\pi\nu}{4h^2}(\liminf_{T\rightarrow\infty}U_T)+\frac{27\pi^2L_x\nu^2}{2^{7/2}h^3}.
\end{aligned}
\end{equation}

Substituting \eqref{evbdext2} into \eqref{evbdext}, we obtain

\begin{equation}
\frac{\nu}{L_xL_yh}\limsup_{T\rightarrow\infty}\average{\left\|\bu(s)\right\|^2}_T\leq\left(\frac{3}{2^{5/2}\pi^2h}+\frac{27\pi}{4}\frac{1}{\Rey}+\frac{27\pi^2}{2^{7/2}}\frac{1}{\Rey^2}\right)\frac{\liminf_{T\rightarrow\infty}U_T^3}{h}.
\end{equation}

\section{The plane Poiseuille flow}
\label{poiseuillesec}
In order to see how sharp our estimates are,
we calculate the characteristic quantities for a specific explicit flow.
It can be easily shown that there exists an explicit solution for the stationary version of the channel flow problem in this geometry, known as the
plane Poiseuille flow:

\begin{equation}
\label{poiseuille}
  \bu_{\text{Poiseuille}}(x,y,z)=\frac{P}{2\nu L_{x}}z(h-z)e_{1}.
\end{equation}

A straightforward calculation gives us the following estimates:

\begin{kenergyP}
\label{kenergyP}
We have, for the plane Poiseuille flow,
\begin{equation}
e_{\text{Poiseuille}}=\frac{1}{2L_{x}L_{y}h}\int_{\Omega}\frac{P^2}{4\nu^2
      L_{x}^2}z^2(h-z)^2dxdydz=\frac{h^{4}}{240\nu^{2}L_{x}^{2}}P^{2};
\end{equation}
\begin{equation}
\epsilon_{\text{Poiseuille}}=\frac{\nu}{L_{x}L_{y}h}\int_{\Omega}\frac{P^2}{4\nu^2
   L_{x}^2}(h-2z)^2dxdydz=\frac{h^2}{6\nu L_{x}^{2}}P^{2};
\end{equation}
\begin{equation}
C_{f_{\text{Poiseuille}}}=\frac{12}{Re};
\end{equation}
\begin{equation}
U_{\text{Poiseuille}}=\frac{Ph^{2}}{12\nu L_{x}};
\end{equation}
and
\begin{equation}
\Rey_{\text{Poiseuille}}=\frac{U_{u_P}h}{\nu}=\frac{Ph^{3}}{12\nu^2 L_{x}}.
\end{equation}




\end{kenergyP}

\rmk Notice that the upper bound estimates for the mean
kinetic energy, and mean energy dissipation obtained in the previous sections are sharp in the sense that they are
of the same order (up to a multiplicative constant) as those just presented for
the plane Poiseuille flow.
They are sharp independently of the value of the applied
pressure. As far as we know, these estimates were known
to be sharp only when the applied pressure is low, since in
this case the plane Poiseuille flow is globally asymptotically stable.

\section{The rate of decrease of energy injection with respect to 
  the scales of the flow}
\label{thetransfer}

In the classical theory of homogeneous turbulence, it is argued that for
turbulent flows, the energy injection is concentrated on the large scale 
motions, whereas the energy dissipated into heat due to
the molecular viscosity occurs on scales that are much smaller
than those.

In 1941, Kolmogorov
\cite{kolmogorov} proposed that within a certain range of scales
much lower than the energy injection scales and greater than the energy
dissipative scales, the energy is transferred to the small scales
at a nearly constant rate equal to the energy dissipation rate.
This mechanism is called the energy cascade, and sufficient
conditions were rigorously derived in \cite{fmrt, fmrtcr1,
fmrtcr2} for the existence of this phenomenon. This theory was proposed
in the idealized case of locally homogeneous turbulence, away from
the boundaries, with the injection of energy restricted to the large
scales.
However, it is well known from experiments that for wall bounded turbulence,
   this hypothesis needs to be corrected; see \cite{constantindoering95, TL, Prandtl}.
In particular, the energy injection occurs at arbitrarily small scales.
Our next result gives an upper bound for the rate of decrease 
of mean energy injection at progressively small scales. 

By taking the scalar product of the Navier-Stokes equations with 
the component $\bu_{\kappa',\kappa''}$ of the flow we find
the energy equation for the scales of motion in the range
$[\kappa', \kappa'')$:
\begin{equation}  
  \frac{1}{2}\ddt{} |\bu_{\kappa',\kappa''}|_0^2 + \nu
  \|\bu_{\kappa',\kappa''}\|^2 + b(\bu,\bu, \bu_{\kappa',\kappa''})
    = ((\bbf_P)_{\kappa',\kappa''}, \bu_{\kappa',\kappa''}),
\end{equation}
if $\kappa''<\infty$, and
\begin{equation}  
  \frac{1}{2}\ddt{} |\bu_{\kappa',\infty}|_0^2 + \nu
  \|\bu_{\kappa',\infty}\|^2 + b(\bu,\bu, \bu_{\kappa',\infty}) 
    \leq ((\bbf_P)_{\kappa',\infty}, \bu_{\kappa',\infty}),
\end{equation}
if $\kappa''=\infty$.

The energy injection at each scale associated with 
a wavenumber $\kappa$ due to the pressure gradient is thus given by
\[  \mathfrak{F}_\kappa(\bu)
     = \frac{1}{L_{x}L_{y}h}((\bbf_P)_\kappa,\bu_\kappa).
\]
The energy injection in a range of wavenumbers $[\kappa',\kappa'')$
is given by
\[ \mathfrak{F}_{\kappa',\kappa''}(\bu)
     = \frac{1}{L_{x}L_{y}h}((\bbf_P)_{\kappa',\kappa''},
          \bu_{\kappa',\kappa''}).
\]
In particular, the energy injection into the wavenumbers larger than or
equal to a given wavenumber $\kappa$ is given by 
$\mathfrak{F}_{\kappa,\infty}(\bu)$.

The mean energy injection is given by the average value of those
quantities with respect to a given stationary statistical solution.
In order to estimate the mean energy injection at different length
scales let us prove the following lemma.

\begin{componentestimate}
\label{componentestimate}
The forcing term component $(\bbf_P)_\kappa$, for a given wavenumber $\kappa$,
satisfies
\begin{equation} 
  \left|\left(\mathbf{f}_{P}\right)_{\kappa}\right|_0
     = \begin{cases}
          \displaystyle \frac{2L_y^{1/2}}{L_x^{1/2}h^{1/2}}
          \frac{P}{\kappa},
           &\displaystyle \text{if } \kappa=\frac{k\pi}{h},\; k\in \NN, 
                \; k \text{ odd;} \medskip \\
          \displaystyle
            0 &\displaystyle
              \text{otherwise.}
       \end{cases}
\end{equation}
\end{componentestimate}

\proof First, notice that we can write
\begin{equation}
\label{forcecomp2}
\left(\mathbf{f}_{P}\right)_{\kappa}=\sum_{\lambda_{j,l,k}=\kappa^2}(\mathbf{f}_{P},\mathbf{w}_{j,l,k})\mathbf{w}_{j,l,k},
\end{equation}
so that by the Parseval identity we have
\begin{equation}
\label{Parsexp}
\left|\left(\mathbf{f}_{P}\right)_{\kappa}\right|_0^2=\sum_{\lambda_{j,l,k}=\kappa^2}\left|(\mathbf{f}_{P},\mathbf{w}_{j,l,k})\right|^2.
\end{equation}
We also notice that each projection satisfies
\begin{equation}
\label{projec}
\left(\mathbf{f}_{P},\bw_{j,l,k}\right)=\frac{P}{L_{x}}\int_{\Omega}w_{j,l,k}^1d\bx,
\end{equation}
where  $w_{j,l,k}^1$ denotes the first component of the eigenvector
$\bw_{j,l,k}$. Now, by inspecting the expression \eqref{ev} for $\bw_{j,l,k}$, 
we notice that the integral \eqref{projec} vanishes for
all $(j,l)\neq (0,0)$. Thus, \eqref{Parsexp} reduces to

\begin{equation} \left|\left(\mathbf{f}_{P}\right)_{\kappa}\right|_0^2=P^2L_y^2\sum_{\lambda_{0,0,k}=\kappa^2}\left|\int_{0}^h\hat{w}^1_{0,0,k}(z)dz\right|^2.
\end{equation}

Furthermore, one can deduce from the Stokes problem and the expansion
\eqref{ev} that the component $\hat \bw_{0,0,k}(z)$ satisfies
following one-dimensional eigenvalue problem:
\[ \begin{cases}
    \displaystyle
    -\frac{\partial^{2}\hat{w}_{0,0,k}^1(z)}{\partial z^{2}}
        =\lambda_{0,0,k} \hat{w}_{0,0,k}^1(z), \\
    \displaystyle
    -\frac{\partial^{2}\hat{w}_{0,0,k}^2(z)}{\partial z^{2}}
        =\lambda_{0,0,k} \hat{w}_{0,0,k}^2(z), \\
    \hat{w}_{0,0,k}^2(z)  =0,
\end{cases}
\]
And the normalized solution to this equation is
\[
\hat{\bw}_{0,0,k}(0,0,z)=\frac{1}{L_x^{1/2}L_y^{1/2}h^{1/2}}
      \left(\sin(\frac{k\pi}{h}z), \sin(\frac{k\pi}{h}z),0\right),
\]
with $\lambda_{0,0,k}=(k\pi/h)^2$. 
Then, we can once more reduce \eqref{Parsexp} to
\begin{equation}
  \left|\left(\mathbf{f}_{P}\right)_{\kappa}\right|_0^2
     = \begin{cases}
          \displaystyle
            P^2L_y^2\left|\int_{0}^h\hat{w}^1_{0,0,k}(z)dz\right|^2 &\displaystyle
              \text{if } \kappa=\frac{k\pi}{h}, \text{ for some } k\in \NN,
                \medskip \\
          \displaystyle 0, 
           &\displaystyle
                \text{if } \kappa\neq\frac{k\pi}{h},\text{ for every } k\in\NN
       \end{cases}
\end{equation}

Then the result follows directly from the following calculation:
\begin{equation}
\begin{aligned}
  \int_{0}^h\hat{w}^1_{0,0,k}(z)dz
     &=\frac{1}{L_{x}^{1/2}L_{y}^{1/2}h^{1/2}}\int_{0}^{h}\sin(\frac{k\pi}{h}z)dz\\
     &= \begin{cases}
          \displaystyle
          \left(\frac{2h^{1/2}}{\pi L_{x}^{1/2}L_{y}^{1/2}}\right)\frac{1}{k}, 
                 &\displaystyle \text{if }  k\; \text{is odd}, \medskip \\
          \displaystyle 0,
           &\displaystyle \text{if }  k\; \text{is even}. \qed
       \end{cases}
\end{aligned}
\end{equation}

We now estimate the energy injection at a given wavenumber $\kappa$.

\begin{prop}
  The mean energy injection at a given wavenumber $\kappa$ with 
  respect to an arbitrary stationary statistical solution satisfies
  \begin{equation}
    \average{\mathfrak{F}_\kappa(\bu)}\leq \begin{cases}
          \displaystyle \frac{2}{L_x^{3/2}L_y^{1/2}h^{3/2}}
          \frac{P}{\kappa^2}\|\bu_\kappa\|,
           &\displaystyle \text{if } \kappa=\frac{k\pi}{h},\; k\in \NN, 
                \; k \text{ odd,} \medskip \\
          \displaystyle
            0, &\displaystyle
              \text{otherwise.}
       \end{cases}
  \end{equation}
\end{prop}

\proof We have
\[ \average{\mathfrak{F}_\kappa(\bu)} =
     \frac{1}{L_x L_y h}((\bbf_P)_\kappa,\bu_\kappa)
       \leq \frac{1}{L_x L_y h} |A^{-1/2}(\bbf_P)_\kappa|_0 \|\bu_\kappa\|
       = \frac{1}{L_x L_y h} \frac{1}{\kappa} |(\bbf_P)_\kappa|_0
           \|\bu_\kappa\|,
\]
and the result follows from using Lemma \ref{componentestimate}.
\qed

\begin{upperkappa}
\label{upperkappa}
 The mean energy injection on the modes larger than or equal to a
given $\kappa$ and  with respect to an arbitrary stationary statistical 
solution satisfies
    \begin{equation}
      \average{\mathfrak{F}_{\kappa,\infty}(\bu)} \leq
        \frac{1}{\kappa^{3/2}}\left(\frac{2P}{\pi^{1/2}L_{x}^{3/2}L_y^{1/2}h}\right)
        \average{\left\|\bu_{\kappa,\infty}\right\|}\leq\frac{1}{\kappa^{3/2}}
        \left(\frac{2P}{\pi^{1/2}\nu^{1/2}L_{x}h^{1/2}}\right)\epsilon^{1/2}.
   \end{equation}

\proof We have:
\[ \frac{1}{L_{x}L_{y}h}\average{(\mathbf{f}_{P},\bu_{\kappa,\infty})}
      \leq\frac{1}{L_{x}L_{y}h}|A^{1/2}(\bbf_{P})_{\kappa,\infty}|_0
               \average{\|\bu_{\kappa,\infty}\|}.
\]
Estimating the term
$|A^{-1/2}(\mathbf{f}_{P})_{\kappa,\infty}|_{0}$, 
we have by the Parseval identity that
\[ |A^{-1/2}(\mathbf{f}_{P})_{\kappa,\infty}|_{0}^{2}
     =\sum_{\kappa'=\kappa}^{\infty}
       |A^{-1/2}(\mathbf{f}_P)_{\kappa'}|_{0}^{2}
  =\sum_{\kappa'=\kappa}^{\infty} \frac{1}{{\kappa'}^2}|(\bbf_P)_\kappa'|_0^2.
\]
Then, if $k$ is the smallest odd number such that
$\kappa\leq k\pi/h$, 
and using Lemma \ref{componentestimate}, 
the following estimate holds
\begin{multline*}
   |A^{-1/2}(\mathbf{f}_{P})_{\kappa,\infty}|_{0}^{2}
  = \frac{4L_yh^3P^2}{\pi^4L_x}\sum_{j=k, \;j\;\text{odd}}^\infty \frac{1}{j^4}
  \leq \frac{4L_yh^3P^2}{\pi^4L_x} \int_k^\infty \frac{1}{s^4}ds \\
  = \frac{4L_yh^3P^2}{\pi^4L_x}\frac{1}{3k^3}
  \leq \frac{4L_yP^2}{\pi L_x}\frac{1}{\kappa^3}.
\end{multline*} 
Thus,
\[ \frac{1}{L_{x}L_{y}h}\average{(\mathbf{f}_{P},\bu_{\kappa,\infty})}
     \leq \frac{2P}{\pi^{1/2}L_x^{3/2}L_y^{1/2} h}\frac{1}{\kappa^{3/2}}
       \average{\|\bu_{\kappa,\infty}\|},
\]
which completes the proof.
\qed
\end{upperkappa}

%
%

\end{document}